\documentclass{article}
\usepackage{spconf,amsmath,graphicx}

\usepackage{graphicx}
\usepackage{threeparttable}
\usepackage{colortbl}
\usepackage{bbding}
\usepackage{multirow}
\usepackage{multicol}
\usepackage[table]{xcolor}
\usepackage{makecell}
\usepackage{algorithm}
\usepackage{algorithmic}
\usepackage{color}
\usepackage{xcolor}
\usepackage{amsmath,amssymb}
\usepackage{booktabs}
\definecolor{darkblue}{rgb}{0.0,0.0,0.55}
\definecolor{darkgray}{rgb}{0.66,0.66,0.66}

\title{DDTSE: DISCRIMINATIVE DIFFUSION MODEL FOR TARGET SPEECH EXTRACTION}
\name{\parbox{0.56\linewidth}{Leying Zhang$^{1,2,\dagger}$, Yao Qian$^{2,\ddagger}$, Linfeng Yu$^1$, Heming Wang$^2$ \\
\textit{Hemin Yang$^2$, Shujie Liu$^2$, Long Zhou$^2$, Yanmin Qian$^{1,\ddagger}$}}
\thanks{$^\dagger$ Work done during an internship at Microsoft.}%
\thanks{$^\ddagger$ Corresponding author.}}

\address{$^1$Auditory Cognition and Computational Acoustics Lab \\ MoE Key Lab of Artificial Intelligence, AI Institute\\ Department of Computer Science and Engineering, Shanghai Jiao Tong University, Shanghai, China \\
$^2$Microsoft, USA}
\begin{document}
\maketitle
\ninept
\begin{abstract}
Diffusion models have gained attention in speech enhancement tasks, providing an alternative to conventional discriminative methods. However, research on target speech extraction under multi-speaker noisy conditions remains relatively unexplored. Moreover, the superior quality of diffusion methods typically comes at the cost of slower inference speed. 
In this paper, we introduce the \textbf{D}iscriminative \textbf{D}iffusion model for \textbf{T}arget \textbf{S}peech \textbf{E}xtraction (DDTSE). We apply the same forward process as diffusion models and utilize the reconstruction loss similar to discriminative methods. Furthermore, we devise a two-stage training strategy to emulate the inference process during model training.  DDTSE not only works as a standalone system, but also can further improve the performance of discriminative models without additional retraining. Experimental results demonstrate that DDTSE not only achieves higher perceptual quality but also accelerates the inference process by 3 times compared to the conventional diffusion model. 
\end{abstract}
\begin{keywords}
target speech extraction, speech enhancement, diffusion model, discriminative model
\end{keywords}
\section{Introduction}
\label{sec:intro}
The cocktail party effect, also known as ``selective hearing", is the ability to focus on a single speaker or conversation in a noisy environment~\cite{bronkhorst2000cocktail,zmolikova2023neural,qian2018past}. Target Speech Extraction (TSE) aims to emulate human capability by isolating the clean speech of the target speaker from a noisy mixture. It serves as a valuable tool for enhancing downstream tasks like speech recognition and speaker verification, attracting significant research interests \cite{delcroix2018single,zhang2022enroll,wang2018voicefilter}. 
 
Discriminative and generative models are two different approaches for speech enhancement and target speech extraction tasks. The former learns the best mapping between inputs and outputs, while the latter learns the target distribution, allowing multiple valid estimates~\cite{lemercier2023analysing}. TSE primarily relies on discriminative methods such as DPCCN~\cite{han2022dpccn}, SpEX~\cite{xu2020spex} and Speakerbeam~\cite{vzmolikova2019speakerbeam}. Despite the many advances gained from past research, discriminative methods occasionally show limited generalization abilities towards unseen noise types or speakers~\cite{lemercier2023storm,richter2023speech}. 

Generative methods, particularly diffusion methods, have shown potential in producing natural and diverse speech, thereby attracting significant interest~\cite{yu2024generation,phan2020improving,tai2024dose}. Previous work, such as SGMSE+ and DiffTSE~\cite{ lemercier2023analysing,richter2023speech,welker2022speech,kamo2023target}, has applied score-based diffusion models on speech enhancement and target speech extraction. 
However, the discrepancy between the forward and reverse processes of diffusion models might lead to degradation of model performance~\cite{lay2023reducing}. Moreover, there are few explorations for the target speech extraction task in multi-speaker noisy environments with generative models. Furthermore, enhancing inference efficiency remains a challenge for real-world deployment, due to the need for dozens or even hundreds of inference steps of diffusion models.

In this study, we introduce the \textbf{D}iscriminative \textbf{D}iffusion Model for \textbf{T}arget \textbf{S}peech \textbf{E}xtraction (DDTSE), which combines the forward processes of the diffusion model and the training objective used in the discriminative model. We design a two-stage training method and provide two usage modes.  Our extensive experiments reveal that DDTSE surpasses discriminative methods in perceptual quality, particularly in noisy conditions. Furthermore, when integrated with existing models, denoted as  X+DDTSE, it consistently surpasses standalone discriminative models (i.e., X) and demonstrates potential as an effective plug-in enhancement.  In terms of inference efficiency, DDTSE requires only 10 steps for standalone use and 2 steps for X+DDTSE, respectively.
 The main contributions of the paper can be summarized as follows:

1) We introduce DDTSE, an advanced frequency domain TSE model, utilizing the forward process of the diffusion model and the reconstruction objective of the discriminative model. It is not only applicable in multi- and single-speaker scenarios but also effective in processing environmental noise. 

    2) We design a two-stage training process. It learns to extract the clean speech given to the target speaker embedding in the first stage and aims to bridge the gap between training and inference in the second stage.
    
    3) We provide two usage modes for versatility. DDTSE-only operates as a standalone system for end-to-end TSE, and X+DDTSE rectifies existing discriminative models to enhance overall system performance. Audio samples\footnote{https://vivian556123.github.io/slt2024-ddtse/}are available.

\section{RELATED WORK}
\subsection{Discriminative models}
For an extended period, discriminative methods have been the preferred approach for tasks related to speech enhancement and target speech extraction, such as DPCCN~\cite{han2022dpccn} and Speakerbeam~\cite{vzmolikova2019speakerbeam}. These approaches mainly utilize supervised learning to learn an optimized deterministic mapping between corrupted speech $y$ and the corresponding clean
speech target $x$ as in Fig.\ref{fig:related_work}c.  However, these models may result in unpleasant speech distortions and limit generalization abilities towards unseen noise types or speakers~\cite{lemercier2023storm}. 

\subsection{Diffusion models}
\begin{figure}[htb]
  \centering
\centerline{\includegraphics[width=8.5cm]{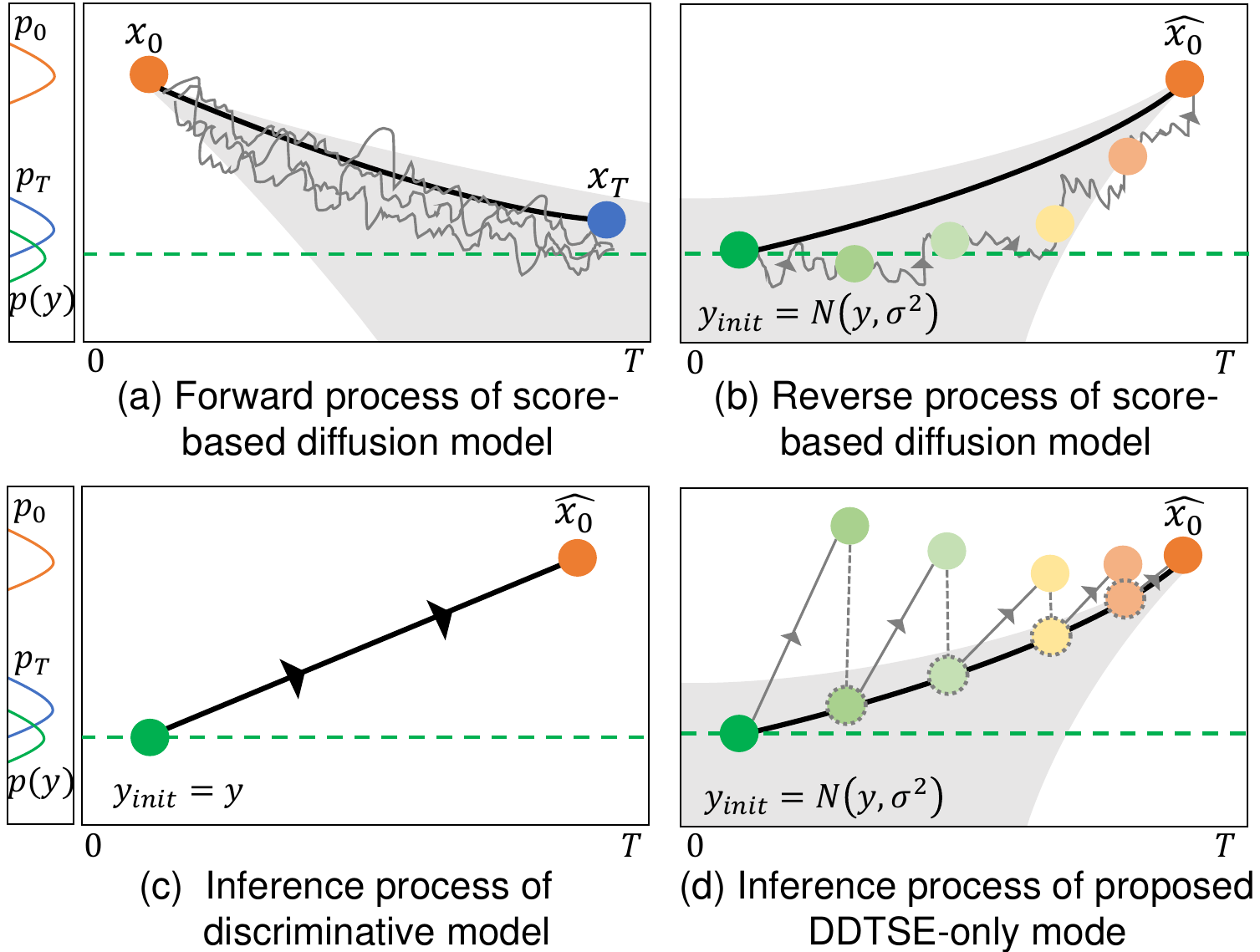}}
\caption{Comparison of score-based diffusion model, discriminative model and our proposed model. The x-axis represents the timestep. (a) and (b) are the forward and reverse process of score-based diffusion model~\cite{ lemercier2023storm,richter2023speech}. (c) is the inference process of discriminative method with one-step prediction. (d) is the inference process of our proposed DDTSE-only mode. The solid gray line is the model prediction in each step. The dashed gray line and the dotted circles are the results obtained by adding noise according to Eq.\ref{eq:mu} and \ref{eq:sigma}.}
\label{fig:related_work}
\end{figure}

Recently,  there are various diffusion models, especially score-based diffusion models, designed for speech enhancement and target speech extraction tasks~\cite{ lemercier2023analysing,richter2023speech,tai2024dose,welker2022speech, kamo2023target,yen2023cold, nguyen2023conditional}. 
The forward process is defined through a linear stochastic differential equation (SDE) and gradually turns data into noise. The reverse process is to sample a target data point from Gaussian noise and invert this process with a reverse-time SDE. The forward and reverse process of score-based diffusion methods is shown in Fig.\ref{fig:related_work}a and b.

As introduced in \cite{richter2023speech, nguyen2023conditional}, the forward process is modeled as the solution to an SDE as in Eq.\ref{eq:forward_sgmse}, where $\mathbf{y}$ is the spectrogram of corrupted speech, $\mathbf{x}_0$ is the spectrogram of target clean speech, $\mathbf{x}_t$ is the state of the process at time $t \in [0,T]$, $\mathbf{f}$ and $g$ are drift and diffusion coefficient function parameterized by $\gamma, \sigma_{\max}, \sigma{\min}$. $\mathbf{w}$ is the standard Wiener process.   
\begin{equation}
\label{eq:forward_sgmse}
    \mathrm{d} \mathbf{x}_t=\underbrace{\gamma\left(\mathbf{y}-\mathbf{x}_t\right)}_{\mathbf{f}\left(\mathbf{x}_t, \mathbf{y}\right)} \mathrm{d} t+\underbrace{\left[\sigma_{\min}\left(\frac{\sigma_{\max}}{\sigma_{\min}}\right)^t \sqrt{2 \log \left(\frac{\sigma_{\max}}{\sigma_{\min}}\right)}\right]}_{g(t)} \mathrm{d} \mathbf{w}
\end{equation}
Eq.\ref{eq:forward_sgmse} describes a Gaussian process, the mean and variance of ${\mathbf{x}_t}_{t\in[0,T]}$ can be derived as in Eq.\ref{eq:mu} and \ref{eq:sigma},  when its initial conditions are known~\cite{sarkka2019applied}. The solution for $\mathbf{x}_t$, called perturbation kernel, is shown in Eq.\ref{eq:forward}. In practice, we sample each $\mathbf{x}_t$ through Eq.\ref{eq:sample_xt}, and the random noise is $\mathbf{z} \sim \mathcal{N}(0, \mathbf{I})$. 
\begin{equation}
\label{eq:mu}
\mu(\mathbf{x}_0,\mathbf{y},t) = \exp^{-\gamma t}\mathbf{x}_0 + (1-\exp^{-\gamma t} )\mathbf{y}
\end{equation}
\begin{equation}
\label{eq:sigma}
\sigma(t)^2=\frac{\sigma_{\min }^2\left(\left(\sigma_{\max } / \sigma_{\min }\right)^{2 t}-\exp^{-2 \gamma t}\right) \log \left(\sigma_{\max } / \sigma_{\min }\right)}{\gamma+\log \left(\sigma_{\max } / \sigma_{\min }\right)}
\end{equation}
\begin{equation}
\label{eq:forward}
q(\mathbf{x}_{t} |\mathbf{x}_0, \mathbf{y}) = \mathcal{N}(\mathbf{x}_{t}; \mu(\mathbf{x}_0,\mathbf{y},t), \sigma(t)^2)
\end{equation}
\begin{equation}
\label{eq:sample_xt}
    \mathbf{x}_t = \mu(x_0,\mathbf{y}, t) + \sigma(t)\mathbf{z}
\end{equation}
For each SDE in the form of Eq.\ref{eq:forward_sgmse}, the corresponding reverse-time SDE is defined by Eq.\ref{eq:reverse_gradient}, where $\mathrm{d}t$ is a negative timestep in the reverse process~\cite{anderson1982reverse}. We can train a neural network to approximate  $\nabla_{\mathbf{x}_t} \log p_t(\mathbf{x}_{t}|\mathbf{y})$, which is the score of the perturbation kernel. According to \cite{song2020score,vincent2011connection}, the loss function takes the form in Eq.\ref{eq:loss_sgmse}.
\begin{equation}
\label{eq:reverse_gradient}
\mathrm{d} \mathbf{x}_t = \left[f\left(\mathbf{x}_{\mathbf{t}},\mathbf{y}\right)-g(t)^2 \nabla_{\mathbf{x}_t} \log p_t(\mathbf{x}_{t}|\mathbf{y})\right] \mathrm{d} t+g(t) \mathrm{d} \mathbf{w}
\end{equation}
\begin{equation}
\label{eq:reverse}
\mathrm{d} \mathbf{x}_t \approx\left[f\left(\mathbf{x}_{\mathbf{t}},\mathbf{y}\right)-g(t)^2 s_\theta\left(\mathbf{x}_t, \mathbf{y}, t\right)\right] \mathrm{d} t+g(t) \mathrm{d} \mathbf{w}
\end{equation}
\begin{equation}
\label{eq:loss_sgmse}
\underset{\theta}{\min } \mathbb{E}_{(\mathbf{x}_{0},\mathbf{x}_{t})\sim q(x_0)q(\mathbf{x}_{t} |\mathbf{x}_0, \mathbf{y}), s,t} \left[\left\|\mathbf{s}_\theta\left(\mathbf{x}_t, \mathbf{y}, t\right)+\frac{\mathbf{z}}{\sigma(t)}\right\|_2^2\right]
\end{equation}
These diffusion methods can generate natural speech, however, as shown in Fig.\ref{fig:related_work}a and b, there exists a discrepancy between the terminating distribution of the forward process (the distribution of $p_T$ for $x$) and the prior used for solving the reverse process at inference (the distribution of $p(y)$)~\cite{lay2023reducing}. This is mainly because of the exponential characteristic of Eq.\ref{eq:mu} that $x_T = \mu(x_0, \mathbf{y}, T) \neq \mathbf{y}$. Moreover, diffusion models require many inference steps to achieve good performance, and thus impose computational pressure.  

\subsection{Combination of discriminative and diffusion models}

Given that diffusion and discriminative models each possess advantages and disadvantages, some researchers focus on combining them. At the architectural level, StoRM~\cite{lemercier2023storm} and Diffiner~\cite{sawata2022versatile} leverage pre-processed speech to guide diffusion-based model training. \cite{shi2024diffusion} utilizes generative and discriminative decoders and fuses them. However, they usually require fine-tuning or joint training, and cannot be used as plug-and-play models.
At the objective level, 
WGSL~\cite{ayilo2024diffusion} augments the original diffusion training objective with an L2 
reconstruction loss at each diffusion time-step. To minimize the discrepancy and accelerate the inference process, \cite{lay2024single} proposed a two-stage diffusion training method through BBED SDE~\cite{lay2023reducing}. In the first stage, it uses the generative denoising score matching loss, and in the second training stage it applies the predictive loss. 

Inspired by these approaches, we propose DDTSE, which combines the forward process of the diffusion model and the training objective of the discriminative model. It not only improves the perceptual speech quality of discriminative models, but also achieves 3x speedup compared to the diffusion model for inference. It is applicable for both noisy and clean multi-speaker or single-speaker speech enhancement and target speech extraction. Two inference modes enable DDTSE to be utilized independently or in a rectified manner combined with other discriminative models, thereby offering a comprehensive solution for various scenarios. 

\section{METHODOLOGY}

\subsection{Training Method}
The objective of TSE is to isolate the target speaker's clean speech from a mixture of multiple speakers and ambient noise. Our model processes the mixture $\mathbf{y}$ to retrieve the clean speech $\mathbf{x}_0$. We collect an enrollment speech from the target speaker to extract the speaker embedding $s$ for model conditioning. 

Our model consists of two processes: the forward process and the reverse process. The forward process $q(\mathbf{x}_{t}| \mathbf{x}_0, \mathbf{y})$ is the same as the score-based diffusion model~\cite{richter2023speech}, as shown in Fig.\ref{fig:related_work}a, defined in Eq.\ref{eq:forward}. It gradually turns the clean speech $\mathbf{x}_0$ into the noisy mixture to simulate speech corruption with timestep $t \in [0,T]$. The reverse process is similar to the discriminative method as shown in Fig.\ref{fig:related_work}c, but it transforms speech with Gaussian noise $\mathbf{x}_t$ back to clean speech over timesteps, conditioned on the speaker embedding~$s$. 

Although prior works mainly focus on predicting the score function during  reverse process~\cite{richter2023speech,kamo2023target}, the score-based objective does not properly measure the perceptual quality of the estimated clean speech because it resembles the generative loss typically utilized in unconditional diffusion models rather than supervised speech enhancement tasks~\cite{ayilo2024diffusion}. To address this problem, we train the model $f_\theta$ to predict the clean speech $\mathbf{x}_0$,  conditioned on the speaker embedding~$s$ by minimizing the conditional expectation as in Eq.\ref{eq:loss_ours}:
\begin{equation}
\label{eq:loss_ours}
\min_{\theta} \mathbb{E}_{(\mathbf{x}_{0},\mathbf{x}_{t})\sim q(x_0)q(\mathbf{x}_{t} |\mathbf{x}_0, \mathbf{y}), s,t} \left[\|\mathbf{x}_0 - f_\theta(\mathbf{x}_{t}, s, t)\|^2_2 \right] 
\end{equation}
In order to minimize the discrepancy caused by the exponential decay in the forward process, with this training objective, we divide the training process into two stages. 

\vspace{-0.1cm}
\subsubsection{First training stage}
The first stage enables DDTSE to progressively learn to extract the target speech from the conditioned target speaker embedding. Algorithm \ref{alg:train} shows that at each step of this stage, $\mathbf{x}_t$ is obtained through the forward process defined in Eq.\ref{eq:forward}, and then the model predicts clean speech $\hat{\mathbf{x}_0}_t$.  We denote $\hat{\mathbf{x}_0}_t$ as the predicted clean speech at the $t$th timestep.  $\mu, \sigma$ are defined in Eq.\ref{eq:mu} and Eq.\ref{eq:sigma}, respectively. We incorporate $\lambda(t) > 0$ to control the weight of loss at different timesteps  \cite{ho2020denoising, lu2023understanding}. $d$ measures L2 distance.
\label{sec:mft}

\vspace{-0.1cm}
\subsubsection{Second training stage}
\label{para:t2}
Similar to conventional diffusion models~\cite{lay2023reducing}, when comparing the first training stage (Algorithm \ref{alg:train}) with the inference process (Algorithm~\ref{alg:infer}, to be explained later), we observe two key differences. The first is the discrepancy between the terminating distribution of the forward process, 
 which is $p_T$ shown in Fig.\ref{fig:related_work}a, and the prior used for inference, which is the distribution of $p(y)$ in Fig.\ref{fig:related_work}d. Secondly, as indicated in blue, the substitution of the real value $\mathbf{x}_0$ with the predicted version $\hat{\mathbf{x}_0}_{t+1}$ also causes mismatch.

Hence, we design the second training stage, shown in Algorithm~\ref{alg:mft}, to imitate the inference process during training. This stage incorporates three strategies. Strategy A simulates the first step prediction from $\mathbf{x}_{T} = \mathcal{N}( \mathbf{y}, \sigma(t)^2)$ to $\hat{\mathbf{x_0}}_{T}$, which ameliorates the first discrepancy. Strategy B further emulates the second step prediction, as described in Algorithm \ref{alg:infer}, making our model not only rely on real but also on predicted values at each reverse step. The probabilities, ${p_1}$ and ${p_2}$, of employing these two strategies are increased linearly with the progression of training epochs, as defined by Eq.\ref{eq:probablity}. Strategy C maintains consistency with the first training stage but gradually reduces its probability of sampling.
\begin{equation}
    p_1 = p_2 = \min(0.45, \frac{\text{current training epoch}}{100})
    \label{eq:probablity}
\end{equation}
\vspace{-0.3cm}
\begin{algorithm}
	\caption{First Training Stage}
	\label{alg:train}
	\begin{algorithmic}[1]
		\REPEAT
		\STATE Sample $\mathbf{x}_0,\mathbf{y}$, $t\sim \mathcal{U}[0,1]$, $\mathbf{z} \sim \mathcal{N}( 0, \mathbf{I})$
		\STATE Update $\mathbf{x}_t \leftarrow \mu(\textcolor{cyan}{\mathbf{x}_0}, \mathbf{y},t)+\sigma(t)^2\mathbf{z}$ 
		\STATE Update $\hat{\mathbf{x}_0}_t  \leftarrow f_\theta\left(\mathbf{x}_t,  s, t\right) , \lambda(t)\leftarrow\left(e^t-1\right)^{-1}$
		\STATE Take gradient descent step on $\nabla_\theta ( \lambda(t) d(\hat{\mathbf{x}_0}_t, \mathbf{x}_0))$
		\UNTIL converged
	\end{algorithmic}  
\end{algorithm}
\begin{algorithm}
	\caption{Second Training Stage}
	\label{alg:mft}
\begin{algorithmic}[1]
\REPEAT
\STATE Sample $\mathbf{x}_0,\mathbf{y}$, $t\sim \mathcal{U}[0,1]$, $\mathbf{z} \sim \mathcal{N}(0, \mathbf{I}), p\sim\mathcal{U}[0,1]$
\IF{$p <  p_1$} 
\STATE \textcolor{gray}{\# Strategy A}
\STATE Sample $\mathbf{x}_t\sim \mathcal{N}( \mathbf{y}, \sigma(t)^2)$
\STATE Update $\hat{\mathbf{x}_0}_t \leftarrow f_\theta\left(\mathbf{x}_{t}, s,t\right)$, $ \lambda(t) \leftarrow \left(e^t-1\right)^{-1}$
\ELSIF{$p_1 \leq p < p_1 + p_2 $} 
\STATE \textcolor{gray}{\# Strategy B}
\STATE Sample $\mathbf{x}_t\sim \mathcal{N}( \mathbf{y}, \sigma(t)^2)$
\STATE Update $\hat{\mathbf{x}_0}_t' \leftarrow f_\theta\left(\mathbf{x}_{t}, s,t\right) $ , $ \mathbf{x}_t' \leftarrow \mu(\hat{\mathbf{x}_0}_t', \mathbf{y},t)+\sigma(t)^2\mathbf{z}$ 
\STATE Update $\hat{\mathbf{x}_0}_t \leftarrow f_\theta\left(\mathbf{x}_t', s,t\right)$, $ \lambda(t) \leftarrow \left(e^t-1\right)^{-1}$ 
\ELSE 
\STATE \textcolor{gray}{\# Strategy C}
  \STATE Update $\mathbf{x}_t \leftarrow \mu(\mathbf{x}_0, \mathbf{y},t)+\sigma(t)^2\mathbf{z}$ 
		\STATE Update $\hat{\mathbf{x}_0}_t \leftarrow f_\theta\left(\mathbf{x}_t,  s, t\right) $ , $ \lambda(t) \leftarrow \left(e^t-1\right)^{-1}$
\ENDIF
\STATE Take gradient descent step on $\nabla_\theta ( \lambda(t)d(\hat{\mathbf{x}_0}_t , \mathbf{x}_0))$
\UNTIL converged
	\end{algorithmic}  
\end{algorithm}

\vspace{-0.3cm}
\subsection{Inference Method}
\label{sec:inference}
We hope that our model can not only independently realize the TSE task but also enhance the speech quality beyond the existing TSE model, so we propose two usage modes for inference.

\subsubsection{DDTSE-only}
This mode functions as an end-to-end TSE model, delivering high-quality TSE independently. It combines the one-step prediction of the discriminative model and the randomness of the diffusion model. The inference usage mode of DDTSE-only is illustrated in  Algorithm \ref{alg:infer} and Figure~\ref{fig:related_work}d.  

We start by sampling $\mathbf{x}_T$ from a normal distribution centered on $\mathbf{y}$, and predict target sample $\hat{\mathbf{x}_0}_{T}$. The first step is similar to the one-step generation of discriminative method, and will give a coarse prediction of the clean speech. We believe that this prediction is close to the target clean speech, but it is missing in detail. 

In order to continue rectifying this coarse prediction, similar to training, we introduce the forward process of the diffusion model into the DDTSE inference stage and let the model predict a more accurate target speech than the previous step. At each step, we perform the forward process to add random noise to the prediction results of the previous step $\hat{\mathbf{x}_0}_{t+1}$ conditioned on $\mathbf{y}$. This simulated forward process and the sample after adding noise $\mathbf{x}_t$ are indicated by dashed gray lines and dotted circles in Figure~\ref{fig:related_work}d. Then we predict the clean target sample $\hat{\mathbf{x}_0}_{t}$ from the sample with noise $\mathbf{x}_t$, as shown in the solid gray lines. This procedure repeats $T$ times. Finally, the clean speech is obtained by performing iSTFT on $\hat{\mathbf{x}_0}_0$. 
\begin{algorithm}
	\renewcommand{\algorithmicrequire}{\textbf{Input:}}
	\renewcommand{\algorithmicensure}{\textbf{Output:}}
	\caption{Inference for DDTSE-only}
	\label{alg:infer}
	\begin{algorithmic}[1]
		\STATE Sample $\mathbf{x}_T\sim \mathcal{N}( \mathbf{y}, \sigma(T)^2)$
        \STATE Update $\hat{\mathbf{x}_0}_{T} =f_\theta\left(\mathbf{x}_{T}, s, T\right)$ 
        \FOR{$t = T-1,...,0$}
        \STATE Sample $\mathbf{z} \sim \mathcal{N}( 0, \mathbf{I})$
        \STATE Update $\mathbf{x}_t\leftarrow\mu($\textcolor{cyan}{$\hat{\mathbf{x}_0}_{t+1}$}$, \mathbf{y},t)+\sigma(t)^2\mathbf{z} $
        \STATE Update $ \hat{\mathbf{x}_0}_{t}\leftarrow f_\theta\left(\mathbf{x}_{t}, s,t\right)$ 
\ENDFOR
\RETURN $\text{iSTFT}(\hat{\mathbf{x}_0}_0)$ 
	\end{algorithmic}  
\end{algorithm}

\subsubsection{X+DDTSE}
Based on an existing discriminative TSE model (denoted as X), X+DDTSE can achieve higher system performance and speech quality by rectifying the discriminative model's output $\mathbf{x}_{dis}$. 
Unlike DDTSE-only mode, X+DDTSE substitutes the first step prediction $\hat{\mathbf{x}_0}_T$  in Algorithm \ref{alg:infer} with $\mathbf{x}_{dis}$ in Algorithm \ref{alg:rDDTSE} and conducts the final $N$ steps. Since the X's prediction is quite accurate, the hyper-parameter $N$ is set to a small number, with the corresponding timesteps $t$ close to 0,  in order to avoid introducing much interference in Eq.\ref{eq:mu}. Due to the few steps required, X+DDTSE serves as an efficient speech quality optimizer that can be applied to various discriminative models. Moreover, different from previous work~\cite{lemercier2023storm}, X+DDTSE mode can be directly applied without any fine-tuning for model X, and the required inference steps is reduced from 50 to 2.
\begin{algorithm}
	\caption{Inference for X+DDTSE }
	\label{alg:rDDTSE}
	\begin{algorithmic}[1]
        \STATE $\hat{\mathbf{x}_0}_{T-N+1} = \mathbf{x}_{dis}$ 
        \FOR{$t = T-N,...,0$}
        \STATE Sample $\mathbf{z} \sim \mathcal{N}( 0, \mathbf{I})$
        \STATE Update $\mathbf{x}_t \leftarrow \mu(\hat{\mathbf{x}_0}_{t+1}, \mathbf{y},t)+\sigma(t)^2\mathbf{z}$
        \STATE Update $\hat{\mathbf{x}_0}_{t} \leftarrow f_\theta\left(\mathbf{x}_{t}, s,t\right)$ 
\ENDFOR
\RETURN $\text{iSTFT}(\hat{\mathbf{x}_0}_0)$ 
	\end{algorithmic}  
\end{algorithm}

\vspace{-0.35cm}
\subsubsection{Inference with ensemble}
We repeat the inference process ten times with different random seeds to get various speech signals, and then sum and normalize to get the final waveform, similar to DiffTSE \cite{kamo2023target}. This strategy leverages  randomness and diversity, leading to a more accurate final waveform, as averaging these outputs can reduce anomalies.

\label{sec:ensemble}

 \subsection{Model architecture}

 \begin{figure}[t]
  \centering
\centerline{\includegraphics[width=8.5cm]{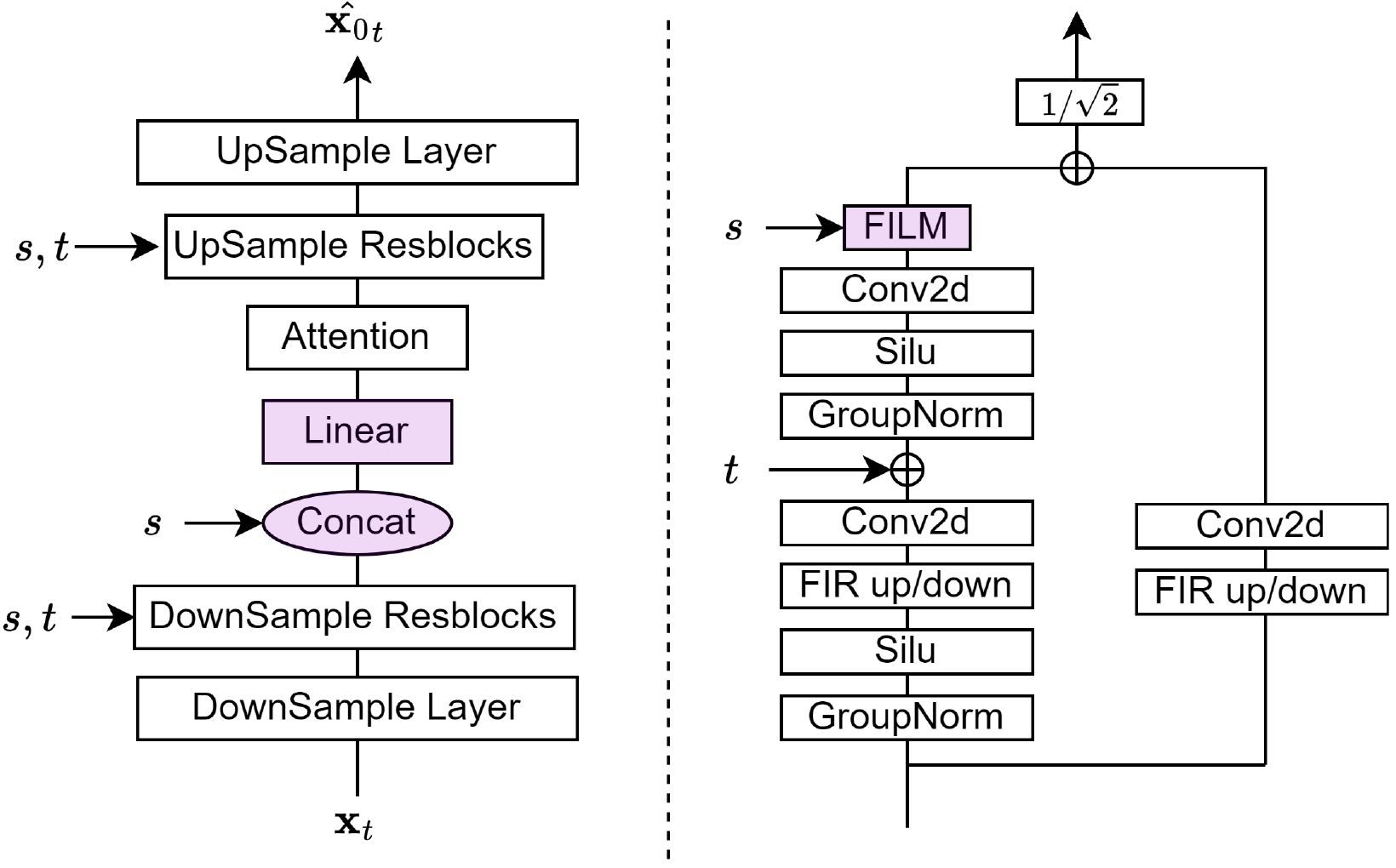}}
\caption{The overall architecture of DDTSE. Left: The model architecture. Right: The (up/down sample) residual block in this model. }
\label{fig:model}
\end{figure}
Fig.\ref{fig:model} depicts the simplified DDTSE model architecture denoted as $f_\theta$ in previous sections. 
It uses a modified NCSN++ network~\cite{song2020score} as the backbone, with modified blocks indicated in purple. The model 
takes speech STFT spectrogram $\mathbf{x}_t$ and a speaker embedding $s$ extracted from a pre-trained speaker verification model as input. The model operates on both real and imaginary parts of the complex spectrogram. We utilize the SiLU activation function \cite{elfwing2018sigmoid}.
We modify its residual block, incorporating the FiLM mechanism of a single linear layer to perceive the target speaker embedding $s$~\cite{perez2018film}. We also concatenate $s$ with the hidden feature within the U-Net, positioned before the self-attention layer to enhance the feature fusion ability.

\begin{table}[t]
\begin{threeparttable}
\centering
\caption{Experimental setup for baselines and our model in standalone usage mode. We abbreviate score estimation loss as S, and clean speech reconstruction loss as R. }
\label{tab:model_config}
\setlength\tabcolsep{4.8pt}
\begin{tabular}{c|c|c|c|c}
\toprule
Scenario                        & Model      & Objective    & Noise & Steps \\ \midrule
\multirow{4}{*}{\shortstack{Multi \\
Speaker \\ Baseline}}  & NCSN++~\cite{song2020score}     & R       & No       & 1     \\
~& DPCCN~\cite{han2022dpccn}      & R       & No       & 1     \\
~& DiffTSE~\cite{kamo2023target}    & S        & Yes      & 30    \\
~& DiffSep~\cite{scheibler2023diffusion}+SV\cite{wang2022wespeaker} & S        & Yes      & 30    \\\midrule
\multirow{3}{*}{\shortstack{Single \\
Speaker \\ Baseline}} & DCCRN~\cite{hu2020dccrn}      & R       & No       & 1     \\
~& SGMSE+~\cite{lemercier2023analysing}    & S        & Yes      & 30    \\
~& WGSL~\cite{ayilo2024diffusion}       & S+R & Yes      & 30    \\ \midrule
Ours  & DDTSE      & R       & Yes      & 10   \\\bottomrule
\end{tabular}
\end{threeparttable}
\end{table}

\section{EXPERIMENTAL SETUP}
\textbf{Data: }We train and evaluate our system on Libri2Mix 16kHz dataset\footnote{https://github.com/JorisCos/LibriMix}~\cite{cosentino2020librimix}, which is derived from LibriSpeech signals~\cite{panayotov2015librispeech} and WHAM noise~\cite{wichern2019wham}. The \texttt{train-360} set is used for training, with \texttt{mix\_both} subset to train multi-speaker model, and \texttt{mix\_single} subset to train single-speaker model. For evaluation, the multi-speaker noisy, multi-speaker clean, single-speaker clean scenarios are \texttt{mix\_both}, \texttt{mix\_clean} and \texttt{mix\_single} test set respectively. The enrollment speech during inference is another speech of the target speaker differing from the target speech.  
All data are transformed into STFT representation with coefficients in \cite{lemercier2023analysing}.


\begin{table*}[t]
\begin{threeparttable}
\centering
\caption{ Performance comparison in multi-speaker noisy and clean scenarios. DDTSE and NCSN++ have the same model architecture. All metrics are the higher the better.
}
\label{tab:big_result}
   \setlength\tabcolsep{5pt}
     \begin{tabular}{l|ccc|cc|c|ccc|cc|c}
    \toprule
\multirow{2}*{Model}  & \multicolumn{6}{c|}{Multi-Speaker Noisy Scenario} &\multicolumn{6}{c}{Multi-Speaker Clean Scenario} \\ 
        ~ & PESQ & ESTOI & SI-SDR & OVRL & DNSMOS & SIM & PESQ &ESTOI & SI-SDR & OVRL & DNSMOS & SIM  \\ \midrule
        Mixture &1.08 & 0.40  & -2.0  & 1.63 & 2.71  & 0.46 & 1.15 & 0.54  & 0.0  & 2.65  & 3.41  & 0.54  \\ \midrule
        DiffTSE\tnote{1} & / & / & / & / & / & / & / & 0.76 & 9.5 & / & / &/   \\ 
        DiffSep+SV & 1.32 &  0.60  & 4.8    & 2.78 &  3.63 & 0.62 & \textbf{1.85} & \textbf{0.79} & 9.6 & 3.14 & \textbf{3.83} & \textbf{0.83}  \\
        DDTSE-only & \textbf{1.60} & \textbf{0.71} & \textbf{7.6} & \textbf{3.28} & \textbf{3.74} & \textbf{0.71} & 1.79 & 0.78 & \textbf{9.9} & \textbf{3.30} & 3.79 & 0.73  \\
        \midrule
        DPCCN & 1.74 & 0.73  & 9.3 & 2.93  & 3.58  & 0.69 &  2.22&0.83  & 13.1  & 3.05 & 3.73 & 0.82  \\ 
        \quad+DDTSE & 1.88 & 0.75 & 9.7  & 3.19 & 3.80 & 0.76 & 2.27& 0.85 & 13.3 & 3.29 & 3.91 & 0.82  \\ 
        \midrule
        NCSN++ & 1.55 & 0.73 & 9.7 & 3.15 & 3.68 & 0.69 & 2.24 & 0.86 & 13.8 & 3.28 & 3.86 & 0.85  \\ 
        \quad+DDTSE & 1.75 & 0.77 & 10.1 & 3.24 & 3.79 & 0.76 & 2.32 & 0.87 & 13.9 & 3.32 & 3.92& 0.85  \\       
  \bottomrule
    \end{tabular}
    \begin{tablenotes}
    \item[1]  Results were reported in ~\cite{kamo2023target}
    \end{tablenotes}
    \end{threeparttable}
\end{table*}
\textbf{Baselines: }Table \ref{tab:model_config} illustrates the model configuration of our proposed DDTSE and the baseline models. In the multi-speaker scenario, we benchmark DDTSE against four baselines: NCSN++, DPCCN, DiffTSE and DiffSep+SV. We train a discriminative model NCSN++~\cite{song2020score}, applying the same architecture as DDTSE. We reproduce DPCCN\footnote{https://github.com/jyhan03/dpccn}~\cite{han2022dpccn}, a widely used discriminative TSE model. We reproduce DiffSep\footnote{https://github.com/fakufaku/diffusion-separation} \cite{scheibler2023diffusion}, a score-based speech separation model, and cascade a speaker verification model~\cite{wang2022wespeaker} after DiffSep for TSE. DiffTSE is a score-based diffusion model for TSE, but it is not accessible for independent reproduction. Our comparative analysis is based on the results reported in the original paper~\cite{kamo2023target}. In single speaker scenario, we benchmark against four reproduced baselines: NCSN++, DCCRN\footnote{https://github.com/asteroid-team/asteroid} \cite{hu2020dccrn} , SGMSE+\footnote{https://github.com/sp-uhh/sgmse}~\cite{lemercier2023analysing}, WGSL~\cite{ayilo2024diffusion}. DCCRN is a conventional discriminative method, while the latter two are the latest score-based diffusion methods for speech enhancement. We re-implement WGSL by ourselves.  All the speaker embedding extractor mentioned in this paper is a ResNet34 speaker verification model pre-trained on VoxCeleb2\footnote{https://github.com/wenet-e2e/wespeaker}~\cite{wang2022wespeaker}. The main distinction between diffusion-based baseline models and proposed DDTSE lies in the training objectives, specifically score-based versus predicting clean data. Compared with discriminative approaches, DDTSE introduce random noise in both the training and inference stages, which brings more randomness and improves the perceptual quality of the generated samples. 

 
\textbf{Settings: }
\label{sec:config}
Parameters defining the forward process in Eq.~\ref{eq:forward_sgmse} is set to $\gamma=1.5,\sigma_{min}=0.05, \sigma_{max}=0.5$. The STFT representation is processed following~\cite{richter2023speech}. We use Adam optimizer and exponential moving average with a decay of 0.999. We use 8 NVIDIA TESLA V100 32GB GPUs for training, with a batch size of 3 samples per GPU. Each sample has 512 STFT frames. We train the first stage with a learning rate of 1e-4 for 500 epochs and the second stage with a learning rate of 5e-5 for 12 epochs. 
DDTSE-only executes 10 inference steps with linearly decreased timesteps from 1 to 0. X+DDTSE performs the last 2 steps out of a total of 10. We select the best-performing model on 20 randomly chosen samples from the dev-set for evaluation.

\textbf{Evaluation metrics: }We evaluate the model performance with both intrusive and non-intrusive speech quality metrics, i.e. with or without clean reference signal \cite{gamper2019intrusive}. Intrusive metrics include Perceptual Evaluation of Speech Quality (PESQ)~\cite{rix2001perceptual}, Extended Short-Time Objective Intelligibility (ESTOI) \cite{jensen2016algorithm}, Scale-invariant Signal-to-Distortion Ratio (SI-SDR)~\cite{le2019sdr}. Non-intrusive metrics, such as  OVRL and  DNSMOS~\cite{reddy2021dnsmos,reddy2022dnsmos}, are used to assess speech quality without clean reference. We use a ResNet34 model pre-trained on VoxCeleb2 to extract speaker embedding for all experiments, and we calculate the cosine speaker similarity (SIM) between speaker embedding of the enhanced speech and the target speech ~\cite{wang2022wespeaker}. All metrics are the higher the better. 

\section{RESULTS AND ANALYSIS}
\subsection{Performance in multi-speaker scenarios}

Table \ref{tab:big_result} shows the performance of DDTSE against both discriminative and generative baselines in scenarios involving multiple speakers under both noisy and clean conditions. 

In the multi-speaker noisy scenario, DDTSE-only outperforms DiffSep+SV on all the metrics. It also demonstrates superior performance on the metrics of OVRL and DNSMOS by comparing with discriminative models, i.e., DPCCN and NCSN++. Notably, the X+DDTSE mode exhibits the highest performance on all the metrics except OVRL. X+DDTSE also improves speaker similarity score (SIM), suggesting a more accurate preservation of individual speaker characteristics during extraction.


In the multi-speaker clean scenario, DDTSE-only surpasses the score-based DiffTSE on the metrics of ESTOI and SI-SDR. Remarkably, DDTSE-only model achieves this by only using a third of the reverse iteration steps, comparing the step number reported in ~\cite{kamo2023target}. This indicates significant enhancements in both signal quality and efficiency. However, it’s observed that the SIM obtained by the DDTSE-only model is the lowest among all the TSE models. This could potentially suggest less discrimination on speaker characteristics when compared to other models, which we will further address in our subsequent work.



\begin{figure}[t]
  \centering
\centerline{\includegraphics[width=8cm]{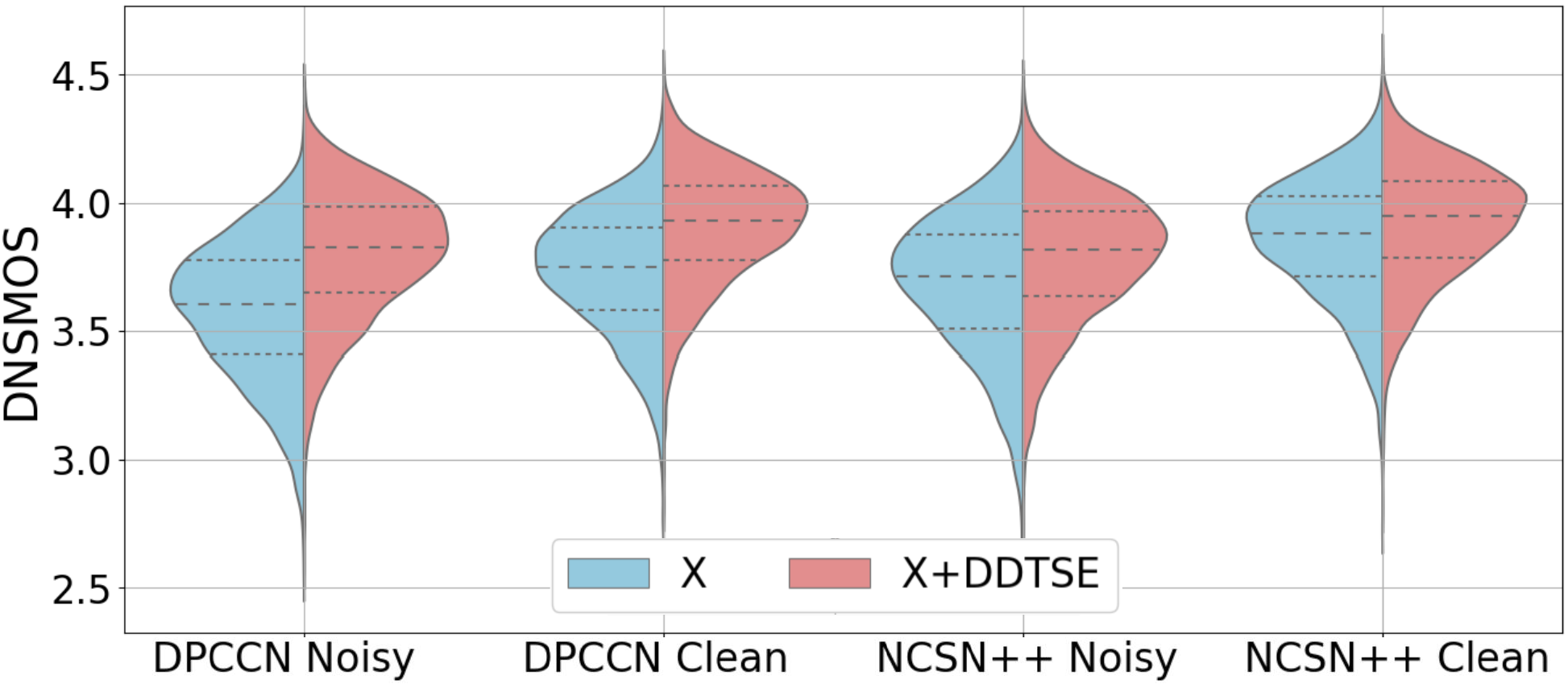}}
\caption{Comparison of DNSMOS distribution between X+DDTSE and corresponding discriminative model (X) DPCCN and NCSN++ in noisy and clean scenarios. Values are the higher the better.}
\label{fig:X+DDTSE-DDTSE}
\end{figure}

Fig.\ref{fig:X+DDTSE-DDTSE} shows the DNSMOS score distributions of various models. We observe that when combined with DPCCN and NCSN++, both models exhibit further improvements in non-intrusive speech quality. Moreover, X+DDTSE mode consistently enhances the performance in both noisy and clean conditions. This suggests that the DDTSE model has potential as a plugin in speech enhancement and extraction tasks.

\begin{table}[t]
\begin{threeparttable}
\centering
\caption{Performance comparison for speech enhancement in single speaker scenario. All metrics are the higher the better. }
\label{tab:enh_result}
   \setlength\tabcolsep{5pt}
    \centering
       \begin{tabular}{l|ccc|cc}
\toprule
        Model & PESQ & ESTOI & SI-SDR & OVRL & DNSMOS  \\ \midrule
       Noisy speech& 1.16   &0.56  &3.5 &1.75 &2.63  \\ \midrule
       NCSN++  & 1.85 & 0.82 & \textbf{12.7} & 3.11 & 3.59 \\
       SGMSE+ & 1.99  &0.82 &11.1 &3.12 &3.60  \\ 
       WGSL & 1.86 & 0.79 & 10.8 & 3.08 & 3.50  \\ 
        DDTSE-only&  \textbf{2.03}  & \textbf{0.83} & 12.6 & \textbf{3.33} & \textbf{3.84}  \\ 
        DDTSE-only\tnote{2} & 2.01 & 0.82 & 12.2 & 3.25 & 3.75  \\ \midrule
      DCCRN  & 2.03 & 0.81 & 13.3 & 2.98 & 3.64  \\ 
        \quad+DDTSE & 2.24   &0.83 &13.7 &3.15 &3.77  \\ 
        \quad+DDTSE\tnote{2} & 2.20 &0.83&13.7 &3.18 &3.80  \\ \bottomrule
    \end{tabular}
 \begin{tablenotes}
        \footnotesize
        \item[2] This model is trained on multi-speaker data  
      \end{tablenotes}
\end{threeparttable}
\end{table}
\subsection{Performance in single-speaker scenario}
Our proposed methods can be generalized to general speech enhancement task. In the single-speaker scenario, speech extraction can be performed without requiring additional enrollment speech. 
We directly extract speaker embedding $s$ from noisy speech $\mathbf{y}$. This is made possible due to the noise robustness of the speaker extractor~\cite{wang2022wespeaker}. The performance comparison in the single-speaker scenario is presented in Table \ref{tab:enh_result}. It indicates that the DDTSE-only model outperforms all other diffusion and discriminative models on all metrics, with the exception of SI-SDR. However, SI-SDR is improved to its highest value when DCCRN is integrated into the X+DDTSE model. This further highlights the potential of DDTSE in enhancing performance when used in conjunction with other models. 
Furthermore, the DDTSE model trained on multi-speaker data achieves comparable performance as the model trained on single-speaker data, indicating the generalization and robustness of DDTSE model. 
We can also employ enrollment speech, as in multi-speaker scenarios, but this only results in a marginal performance gain.




\subsection{Ablation study}
\label{sec:ablation}
Table \ref{tab:ablation} provides an analysis of the individual contributions from the first training stage (T1), the second training stage (T2) and the inference with ensemble strategy (Ensem). These results indicate that with the same total training epochs, by omitting T2 (S1) worsens all metrics, highlighting the necessity for the second training stage. S2 shows that Ensem improves intrusive metrics but slightly reduces non-intrusive quality, suggesting that averaging speech with diversity may introduce undesired distortion. 
Only training with T1 or T2, as shown by S3 and S4, results in a performance decrease on intrusive metrics. Only training with T2 (S4) also causes speaker similarity degradation. 


Moreover, we experimentally find that T2 cannot be trained for too many epochs. Strategy A, introduced in Section \ref{para:t2}, provides one-step coarse prediction of the clean speech at the initial timestep, but it is not precise in detail. Strategy B integrates this prediction into diffusion forward process and deduces a more accurate result. If these strategies are over-presented as training time increases, the resulting estimation errors will lead to sub-optimal performance~\cite{tai2024dose,huang2022prodiff}. Consequently, we limit the training of the second stage to only 12 epochs, as outlined in Section \ref{sec:config}.

\begin{table}[t]
   \begin{threeparttable}
    \centering
    \caption{Ablation Study of DDTSE-only in multi-speaker noisy scenario. T1 and T2 are the first and second training stages. Ensem is the Inference with ensemble strategy.}
    \label{tab:ablation}
    \setlength\tabcolsep{3.6pt}
        \begin{tabular}{c|ccc|cc|cc|c}
    \toprule
        \# & T1 & T2 &Ensem & ESTOI & SI-SDR & OVRL & DNSMOS & SIM  \\ \midrule
    S0 & \Checkmark & \Checkmark & \Checkmark &  0.71 & 7.6 & 3.28 & 3.74 & 0.71  \\ 
        S1 & \Checkmark & \XSolidBrush & \Checkmark & 0.69 & 6.8 & 3.21 & 3.67 & 0.70  \\ 
    S2 & \Checkmark & \Checkmark & \XSolidBrush  & 0.69 & 6.7 & 3.34 & 3.82 & 0.71 \\ 
        S3 & \Checkmark & \XSolidBrush & \XSolidBrush &  0.67 & 5.7 & 3.28 & 3.73  & 0.70 \\ 
        S4 & \XSolidBrush & \Checkmark & \XSolidBrush &  0.66 & 6.1 & 3.27 & 3.79 & 0.61 \\ \bottomrule
    \end{tabular}
    \end{threeparttable}

\end{table}



\begin{table}[]
\begin{threeparttable}
\centering
\caption{Comparison with varying inference steps of DDTSE-only and X+DDTSE modes in multi-speaker noisy scenario. \\}
\label{tab:infer_step}
\setlength\tabcolsep{3pt}
\begin{tabular}{c|c|c|cc|cc|c}
    \toprule
        Model & Steps & RTF & ESTOI & SI-SDR & OVRL & DNSMOS & SIM  \\ \midrule
        \multirow{6}*{\shortstack{DDTSE\\only}}  & 1 & 0.093 & 0.45 & 0.9 & 2.77 & 3.09 & 0.44  \\
        ~ & 5 & 0.273 & 0.64 & 4.9 & 3.20 & 3.59 & 0.66  \\ 
        ~ & 10 & 0.501 & 0.67 & 5.7 & 3.28 & 3.73 & 0.70  \\ 
        ~ & 15 & 0.728 & 0.68 & 5.9 & 3.29 & 3.78 & 0.71  \\ 
        ~ & 20 & 0.954 & 0.68 & 5.9 & 3.28 & 3.80 & 0.71  \\
        ~ & 30 & 1.415 & 0.67 & 5.6 & 3.24 & 3.81 & 0.71   \\ \midrule
        \multirow{4}*{\shortstack{X+\\DDTSE}} & 1 & 0.093 & 0.73 & 9.4 & 3.01 & 3.65 & 0.71   \\ 
        ~ & 2 & 0.139 & 0.75 & 9.4 & 3.18  & 3.81 & 0.76  \\ 
        ~ & 3 & 0.183 & 0.75 & 9.4 & 3.26  & 3.84  & 0.76  \\ 
        ~ & 4 & 0.234 & 0.76 & 9.2 & 3.29 & 3.83 & 0.75  \\  \bottomrule
    \end{tabular}
\end{threeparttable}
\end{table}

\subsection{Inference speed}
Table \ref{tab:infer_step} presents the Real-Time-Factor (RTF), speech quality, and speaker similarity of the generated speech across various inference steps. 
The RTF=$\frac{\text{processing time}}{\text{speech duration}}$ is measured on a single V100 and serves as an efficiency indicator.
We notice that the performance improvements of the DDTSE-only along with the increasing number of steps tend to plateau beyond 10 steps. 
Furthermore, in the X+DDTSE mode, where X means DPCCN, we can achieve robust performance from as few as 2 steps, showcasing rapid processing without sacrificing quality and further eliminating the constraints of slow processing speeds of diffusion models. 
Utilizing our proposed DDTSE inference algorithm, we  achieve significant reductions in the number of inference steps. According to the steps reported in ~\cite{richter2023speech} and ~\cite{lemercier2023storm}, our approaches decrease the steps from 30 to 10 for the DDTSE-only mode, and from 50 to 2 for the X+DDTSE mode, respectively.

\section{CONCLUSION}
We present DDTSE, which combines discriminative training objective and diffusion forward process, designed for target speech extraction and enhancement in multi-speaker and single-speaker scenarios under both noisy and clean conditions. Experimental results demonstrate its effectiveness and efficiency, both as a standalone model and as an additional rectified model.  In the next steps, we will further investigate its potential in other speech generation tasks.

\section{ACKNOWLEDGMENTS}
This work was supported in part by China STI 2030-Major Projects under Grant No. 2021ZD0201500, in part by China NSFC projects under Grants 62122050 and 62071288, and in part by Shanghai Municipal Science and Technology Commission Project under Grant 2021SHZDZX0102.

\bibliographystyle{IEEEbib}
\bibliography{strings,refs}

\end{document}